\documentclass[12pt]{iopart}

\usepackage[dvips]{graphicx}

\begin{document}


\title{Nucleation of interfacial shear cracks in thin films on disordered substrates}

\textbf{}

\textbf{}

\author{M. Zaiser$^1$, P. Moretti$^{1,2}$, A. Konstantinidis$^3$, E.C. Aifantis$^{3,4}$}

\textbf{}

\textbf{}

\address{$^1$The University of Edinburgh, Center for Materials Science and Engineering, The King's Buildings, Sanderson Building, Edinburgh EH93JL, UK}

\address{$^2$Departament de Fisica Fonamental, Facultat de Fisica, Universitat de Barcelona, Marti i Franques 1, E-08028 Barcelona, Spain}

\address{$^3$Laboratory of Mechanics and Materials, Aristotle University of Thessaloniki, 54124 Thessaloniki, Greece}

\address{$^4$Center for the Mechanics of Material Instabilities and Manufacturing Processes (MMIMP), Michigan Tech., Houghton, USA}

\eads{\mailto{M.Zaiser@ed.ac.uk}, \mailto{paolo.moretti@ub.edu},
  \mailto{akonsta@gen.auth.gr }, \mailto{mom@mom.gen.auth.gr}}
\begin{abstract}

We formulate a theoretical model of the shear failure of a thin film tethered to a rigid substrate. The interface between film and substrate is modeled as a cohesive layer with randomly fluctuating shear strength/fracture energy. We demonstrate that, on scales large compared with the film thickness, the internal shear stresses acting on the interface can be approximated by a second-order gradient of the shear displacement across the interface. The model is used to study one-dimensional shear cracks, for which we evaluate the stress-dependent probability of nucleation of a critical crack. This is used to determine the  interfacial shear strength as a function of film geometry and statistical properties of the interface. 
\end{abstract}

\section{Introduction}

In this study we investigate the interfacial failure of thin films subject to shear loads. We consider situations where the interface with the substrate is disordered, such that the interfacial shear strength and fracture energy are random functions of position. While most of the mechanics literature has focused on the properties of interface cracks in systems where the properties of the interface are spatially homogeneous, failure of disordered interfaces with randomly fluctuating strength has been investigated extensively in the statistical physics community (see e.g. \cite{alava06,raischel05,hidalgo02,arndt01}). Besides obvious applications to materials problems such as shear failure of coatings and shear-induced delamination of thin films, the problem at hand has some interesting applications in geosciences \cite{zaiser04,fyffe04a,fyffe04b}, where models similar to the one studied in this paper have been applied to the initiation of snow avalanches and landslides. 

In the present study we investigate theoretically how random variations of interface toughness affect the nucleation of interface cracks. We assume a one-dimensional geometry and evaluate the crack nucleation probability as a function of stress, geometry, and the statistical parameters characterizing the interfacial strength distribution. The theoretical arguments are validated by comparison with numerical simulations.

\section{Formulation of the model}

\begin{figure}
  \centering
  \includegraphics{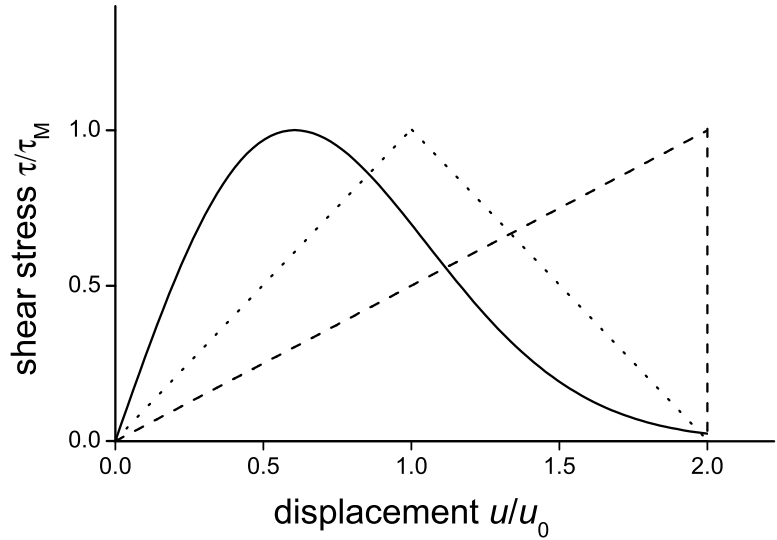}
  \caption{Shear strength versus interfacial displacement (schematically); full line: semi-brittle behaviour, dashed line:
  brittle behaviour; dotted line: piecewise linear approximation to the semi-brittle case as used in our simulations.}
  \label{fig:tau}
\end{figure}

We consider a thin elastic film tethered to a rigid substrate. The interface with the substrate is modeled as a cohesive layer in the plane $z=0$. The response of the interface to shear loads is characterized by a scalar stress-displacement relationship $\sigma _{xy} (x,z=0)=\tau (u)$ where $\sigma _{xy} $ is the shear stress at the interface, $\tau $ is the interfacial shear strength, and  $u(x)=w_{x} (x,z=0)$ is the shear displacement across the interface.  The film is loaded in plane shear by spatially homogeneous tractions applied to its free surface, giving rise to a space-independent 'external' shear stresss $\sigma _{xy}^{{\rm EXT}}$.  The maximum stress that can be supported by the interface is denoted as $\tau _{M} $, and the failure energy is given by the integral
\begin{equation}
w_{f} =\int \tau (u){\rm d}u=:\tau _{{\rm M}} u_{0} 
\label{wf}
\end{equation}
where $u_0$ denotes the characteristic displacement-to-failure. Stress-displacement relationships are schematically shown in Figure \ref{fig:tau} for the semi-brittle (full line) and brittle cases (dashed line). 
Structural disorder is introduced into the model in terms of statistical variations of $\tau _{M}$ which is considered a stochastic process with prescribed statistical properties that will be specified below. 

To analyze crack nucleation and propagation, we have to evaluate the internal stresses associated with the interfacial displacement field $u(x)$. The elastic energy functional associated with a generic displacement vector field $\bi{w}(\bi{r})$ is given by
\begin{equation}
H(\bi{w})=\frac{\mu }{2} \int \left[\alpha ({\rm div} \,\bi{w})^{2} +({\rm curl}\,\bi{w})^{2} \right] \; {\rm d}^{3} r
\label{Hw}
\end{equation}
where $\alpha =(2-2\nu )/(1-2\nu )$, and $\nu$ is Poisson's number. Energy minimization gives the equilibrium equation
\begin{equation}
\nabla ^{2} {\bi{w}}+\frac{1}{1-2\nu } {\rm grad} \left({\rm div}\, {\bi{w}}\right)=0
\label{equil}
\end{equation}

We solve Eqn. \ref{equil} in Fourier space, imposing along the $z=0$ plane the boundary conditions $w_{x} (x,z=0)=u(x),\; w_{y} =w_{z} =0$ and making the crucial assumption that the film thickness $D$ is much smaller than the characteristic length of variations in the displacement field: $\left|kD\right|<<1$ where $k$ is the wave vector modulus of any characteristic variations in $\bi{w}$. The elastic energy can then be written as
\begin{equation}
H(u)=\frac{1+\alpha }{2} D\mu \int \left(\frac{\partial u}{\partial x} \right) ^{2} {\rm d}x
\label{Hu}
\end{equation}

The total energy of the system is obtained by adding to the elastic energy \textit{H} the work done by the shear stresses at the interface: 
\begin{equation}
G(u)=-\int {\rm d}x\left[\int _{0}^{u(x)}\left(\sigma _{xy}^{{\rm EXT}} -\tau (u)\right)\; {\rm d}u \right] 
\label{Gu}
\end{equation}

Minimizing the total energy functional $E(u) = G(u) + H(u)$ leads to the equilibrium condition 
\begin{equation}
I\frac{\partial ^{2} u}{\partial x^{2} } +\sigma _{xy}^{{\rm EXT}} -\tau (u)\le 0
\label{equil_u}
\end{equation}
where the gradient coefficient is given by $I = 2D\mu(1+\alpha)$.  Equations of this type have been studied by Aifantis and co-workers in the context of shear and slip bands in metal plasticity (see e.g. \cite{zbib88}). It may be noted that in these cases the constitutive relation (\ref{equil_u}) contains a strain variable (shear strain or equivalent strain) in place of the displacement variable $u$. The mathematical structure is, however, the same as in the present problem.

\section{Energy of a shear crack and critical crack size}

We first consider interfaces with space independent fracture toughness. This serves to derive some relations and introduce notations that will prove useful in our subsequent analysis of the disordered system. We assume that a mode-II crack of finite length exists along the interface. Such a crack is characterized by a displacement field which, for $x \to \infty$ , starts from a value $u_{0}$ on the left stable branch of the $\tau(u)$ curve, goes through a maximum $u_{1}$ which without loss of generality we assume at $x=0$, and then reverts to $u_{0}$ for $x \to -\infty$. The derivation of the corresponding solutions of Eqn. (\ref{equil_u}) has been discussed elsewhere \cite{fyffe04b}. In the limit of long cracks (small stresses) we find that the displacement profile is approximately parabolic,
\begin{equation}
u(x)=\frac{\left(l^{2} -x^{2} \right)\sigma _{xy}^{{\rm EXT}} }{2I}
\label{wf}
\end{equation}
where the parameter $l$ defines the crack length. This equation describes the displacement profile as long as $u \gg u_0$ such that the shear strength $\tau(u) \approx 0$ along the length of the crack. The parabolic profile is complemented by two boundary layers where the shear strength $\tau(u)$ goes through the curve depicted in Figure \ref{fig:tau}. For a long crack the contribution of these boundary layers can be neglected, and the energy is approximately given by
\begin{equation}
E(l) \approx -\frac{l^{3}}{3I} \left(\sigma _{xy}^{{\rm EXT}} \right)^{2} +2w_{f} l
\label{El}
\end{equation}

This energy has a saddle point at the critical crack length $l_{\rm c} =\sqrt{2w_{\rm f} I}  / \sigma _{xy}^{\rm EXT}$. Cracks above this length are unstable under the load $\sigma _{xy}^{\rm EXT}$ and lead to interface failure. The energy required to create a critical crack is $E_{\rm c} = 4 w_{\rm f} l_{\rm c}/3$, i.e., it decreases in inverse proportion with the applied stress. However, the barrier is always finite and, as emphasized by Arndt and Nattermann \cite{arndt01} for the case of bulk cracks, at physically meaningful stress levels the typical crack nucleation energies are many orders of magnitude above typical thermal energies. Hence, crack nucleation cannot occur as a result of thermally activated processes. However, energy barriers may be substantially reduced or even eliminated at certain locations due to the influence of structural disorder. In the following we analyse the conditions under which crack nucleation may occur spontaneously in a disordered medium. 

\section{Shear crack nucleation at a disordered interface}

A disordered interface can be described by a randomly varying peak shear strength $\tau_{\rm M}(x)$ or, equivalently, failure energy $w_{\rm f}(x)$. In this case, the energy expression, Eqn. (\ref{El}), modifies to
\begin{equation}
E(l)\approx -\frac{l^{3} }{3I} \left(\sigma _{xy}^{\rm EXT} \right)^{2} +\int _{-l}^{l}w_{f}  (x){\rm d}x=\int _{-l}^{l}\left[w_{f} (x)-F(x)\right] \; {\rm d}x
\label{El1}
\end{equation}
where $F(x)=\left(x\sigma _{xy}^{{\rm EXT}} \right)^{2} /(2I)$ is the effective driving force acting on the edge of a crack of length \textit{x}. 

A sufficient criterion for spontaneous nucleation of a propagating shear crack starting from the position $x=0$ is given by $(\partial E/\partial l)<0\; \forall l$ or, equivalently, by $F(l)>w_{f} (l)\; \forall l$. To estimate the probability for this to happen, we consider the case of short-range correlated disorder where the failure energy variations can be characterized by a finite correlation length $\xi$. We split the interface into segments of length $\xi$ and treat the shear strengths in each segment as independent, identically distributed random variables with probability distribution $P(w_{\rm f})$.
The condition that the crack can advance across the $n$th segment is $F(n\xi )>w_{\rm f} (n\xi)$, the probability for this is $P(F(n\xi))$, and the crack nucleation probability is $W_{\rm nucl} = \prod_nP(F(n\xi))$. Taking the logarithm and reverting to continuous variables, we obtain
\begin{equation}
\ln W_{\rm nucl} =\frac{1}{\xi} \int \ln P(F(x)){\rm d}x 
\label{Wn1}
\end{equation}

To evaluate this probability we have to specify the probability distribution characterizing the variations of interfacial strength. We assume a Weibull distribution with characteristic failure energy $w_0$ and modulus $\beta$:
\begin{equation}
P(w_{{\rm f}} )=1-\exp \left[-\left(\frac{w_{{\rm f}} }{w_{0} } \right)^{\beta } \right]
\label{Pwf}
\end{equation}
The mean failure strength is then given by $\langle w_{\rm f} \rangle =  w_0 \Gamma(1+1/\beta)$ where $\Gamma(x)$ denotes the Gamma function, and the coefficient of variation (COV = variance/mean) of the distribution is calculated as
${\rm COV}=[\Gamma(1+2/\beta)/(\Gamma(1+1/\beta))^2-1]^{1/2}$.

Using the driving force given above, the crack nucleation probability is then evaluated as
\begin{eqnarray}
\ln W_{\rm nucl}&=& \frac{1}{\xi} \int \ln P(F(x)){\rm d}x \nonumber \\
&=& \frac{\sqrt{2I\langle w_{\rm f}\rangle}}{\xi \sigma _{xy}^{\rm EXT}} \frac{1}{\Gamma(1+1/\beta)} \int \ln [1-\exp (-s^{2\beta} )]{\rm d}s = -\frac{\sqrt{2I\langle w_{\rm f}\rangle}}{\xi \sigma _{xy}^{\rm EXT}} g(\beta)\nonumber\\
\label{Wn2}
\end{eqnarray}
where $s=x/l_{\rm c}$ is the ratio between the crack size and the size of a deterministic critical crack. Hence we obtain that
\begin{equation}
W_{\rm nucl} =\exp \left[-g(\beta)\frac{\sqrt{2I\langle w_{\rm f}\rangle}}{\xi \sigma_{xy}^{\rm EXT}} \right]= \exp \left[-\frac{E_{{\rm C}} }{k_{{\rm B}} T_{{\rm eff}}} \right].
\label{Wn3}
\end{equation}

This has the structure of an Arrhenius equation where the activation energy is the energy of a critical crack in the system without disorder, and the place of temperature is taken by an effective `disorder temperature' $k_{\rm B} T_{\rm  eff} = 4 \xi \langle w_{\rm f}\rangle/3g(\beta)$. For large $\beta$, we may approximate the integrand in Eqn. (\ref{Wn2}) by $\ln s^{2 \beta}$ for $s<1$, and 0 for $s>1$. In physical terms, this means that only crack sizes up to the critical one significantly affect the nucleation probability, which is thus tantamount to the probability of forming a critical `deterministic' crack. In this limit, we simply obtain that $g(\beta) = 2\beta$. With this approximation, the effective `disorder temperature' is directly proportional to the variance of the Weibull distribution times the correlation length of the strength variations.  

Nucleation of a propagating crack implies system failure. However, Eqn. (\ref{Wn3}) cannot be directly interpreted as a system failure probability: While Eqn. (\ref{Wn3}) is evaluated under the assumption that the crack starts from the origin, nucleation from any other position would produce a similar result. There are $N=L/\xi$ potential nucleation sites in a system of length $L$, and system failure occurs when a crack nucleates at the weakest of these. To evaluate the failure stress distribution, we use an argument from extreme order statistics. Interpreted as a function of stress, $W_{\rm nucl}(\sigma _{xy}^{{\rm EXT}})$ gives the probability that the system has, at any stress larger than $\sigma _{xy}^{{\rm EXT}}$, produced a propagating crack starting from a given nucleation site. $1-W_{\rm nucl}(\sigma _{xy}^{{\rm EXT}})$ is the probability that this has not yet happened, and $[1-W_{\rm nucl}(\sigma _{xy}^{{\rm EXT}})]^N$ is the probability that the system as a whole is still intact at the stress $\sigma _{xy}^{{\rm EXT}}$. Hence we write the system failure probability as
\begin{equation}
P(\sigma _{xy}^{{\rm EXT}}) = 1 - \left[1-W_{\rm nucl}(\sigma _{xy}^{{\rm EXT}})\right]^N
\label{Pfail1}
\end{equation}
or
\begin{equation}
P(\sigma _{xy}^{{\rm EXT}}) = 1 - \exp\left[N \ln(1-W_{\rm nucl})\right]
\label{Pfail2}
\end{equation}
For large systems, the nucleation probability at each site will be quite small, and we can therefore approximate
$\ln(1-W_{\rm nucl}) \approx W_{\rm nucl}$ to obtain (up to logarithmic corrections in the exponent)
\begin{equation}
P(\sigma _{xy}^{{\rm EXT}}) = 1 - \exp\left[\exp\left( -g(\beta)\frac{\sqrt{2I \langle w_{\rm f}\rangle}}{\xi \sigma _{xy}^{\rm EXT}} + \ln(N)\right) \right]
\label{Pfail3}
\end{equation}
This expression can be interpreted as the distribution of system failure stresses. Reciprocal failure stresses are Gumbel distributed, and the characteristic failure stress (e.g. the median of the failure stress distribution) scales like
\begin{equation}
\sigma _{xy}^{\rm EXT} \propto g(\beta )\frac{\sqrt{2I \langle w_{\rm f}\rangle}}{\xi \ln(N)}.
\label{sigfail}
\end{equation}
This stress decreases logarithmically with system size. It is interesting to compare this relationship with the deterministic failure stress of a system containing a crack of length $l$: $\sigma _{xy}^{{\rm EXT}} =\sqrt{2w_{\rm f} I}/l$. It is evident that, in Eqn. (\ref{sigfail}),  the correlation length of the disorder plays a very similar role to the crack size in case of a interface without disorder.

\section{Comparison with simulation results}

To test the above theoretical calculations, we use a lattice automaton technique where we evaluate the displacements at discrete sites $x_i$ on a one-dimensional lattice with lattice constant $\xi$. Accordingly, we replace $u_{xx}$ in Eqn. (\ref{equil_u}) by the corresponding discrete second-order gradient. Furthermore we approximate the strength-displacement characteristics by a piecewise linear curve as shown by the dotted line in Figure \ref{fig:tau}. Nondimensional variables are defined through 
\begin{equation}
T=\frac{\tau }{\left\langle \tau _{{\rm M}} \right\rangle } \; ,\; S=\frac{\sigma _{xy}^{{\rm EXT}} }{\left\langle \tau _{{\rm M}} \right\rangle } =U=\frac{u}{u_{0} } \; ,\; X=\frac{x}{\xi} 
\label{scale}
\end{equation}
such that the average peak strength and fracture energy are by definition equal to 1. In these coordinates the equilibrium equation reads
\begin{equation}
J\frac{\partial ^{2} U}{\partial X^{2} } + \Sigma - T \le 0
\label{equil_U}
\end{equation}
where the interaction constant $J$ is given by $J = (Iu_0)/(\tau_{\rm M} \xi^2)$. The length of a critical crack in a homogeneous system is in non-dimensional coordinates given by $L_{\rm c} = \sqrt{2 J}/\Sigma$, and the failure strength distribution of the disordered system, Eq. (\ref{Pfail3}), reads 
\begin{eqnarray}
P(\Sigma) &=& 1 - \exp\left[\exp\left(- g(\beta) L_{\rm c}(\Sigma) + \ln(N)\right) \right]\nonumber\\
&=& 1 - \exp\left[\exp\left(- \frac{g(\beta) \sqrt{2 J}}{\Sigma} + \ln(N)\right) \right] .
\label{Pfailscaled}
\end{eqnarray}

In our simulations, random values for the local peak strengths are generated as Weibull distributed random fields with average 1, spatial correlation length 1 and Weibull parameter $\beta$. Note that the nondimensional value of the parameter $\beta$, together with the interaction constant $J$ and the system size $N$, are the only independent parameters of the problem.

The system is slowly loaded by increasing the external stress $\Sigma$ from zero in small steps until sites become unstable as the local (external plus internal) stress exceeds the local shear strength. The displacement at all unstable sites is then increased by a small amount $\Delta U$. New internal stresses are computed for all sites and it is checked again where the sum of external and internal stresses exceeds the local strength. The displacement at the now unstable sites is further increased, and the process is repeated until the system has reached a new stable configuration. Then the external stress is increased again and so on until the system has failed completely ($U > 2$ for all sites). The corresponding critical stress is recorded, and the procedure is repeated for different realizations. 

\begin{figure}
  \centering
  \includegraphics{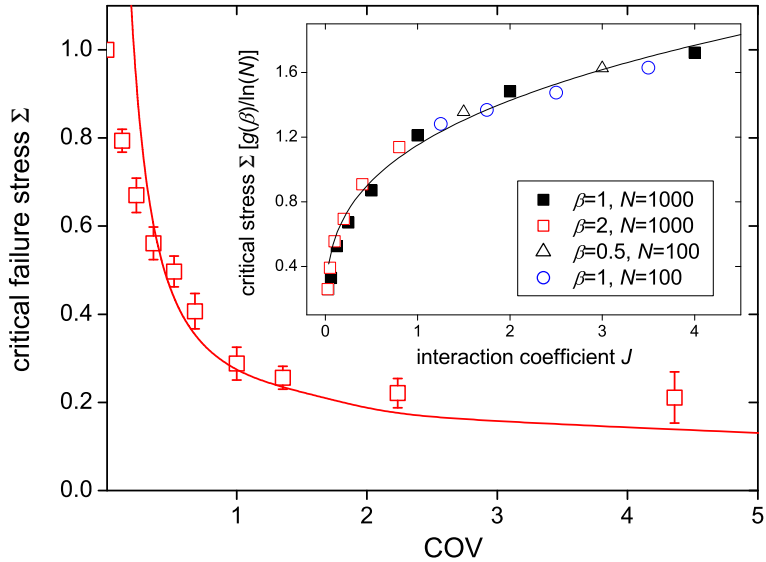}
  \caption{Main graph: failure stress distribution, parameters: $J = \beta = 1, N=100$, 1000 simulations; full line: theoretical
  fit using Eq. (\ref{Pfailscaled}); the fit yields $g(\beta)\sqrt{2J} = 3.00$. Inset: Size dependence of the mean failure
  stress, parameters $J=\beta=1$; full line: $\Sigma = 3/(\ln(N)+3)$.}
  \label{fig:dist}
\end{figure}

Figure \ref{fig:dist} shows a failure stress distribution determined for $J=\beta=1$ in a system of size $N=100$. Fitting the functional form of Eq. (\ref{Pfailscaled}) by using $g(\beta)\sqrt{2J}$ as a fit parameter yields $g(\beta) = 2.12$ as compared to $g(\beta) = 1.64$ obtained by direct computation. The discrepancy is acceptable in view of the approximations made in the derivation of the failure stress distribution. Using the same parameter, the size dependence of the mean failure stress is well reproduced (inset in Figure \ref{fig:dist}).

\begin{figure}
  \centering
  \includegraphics{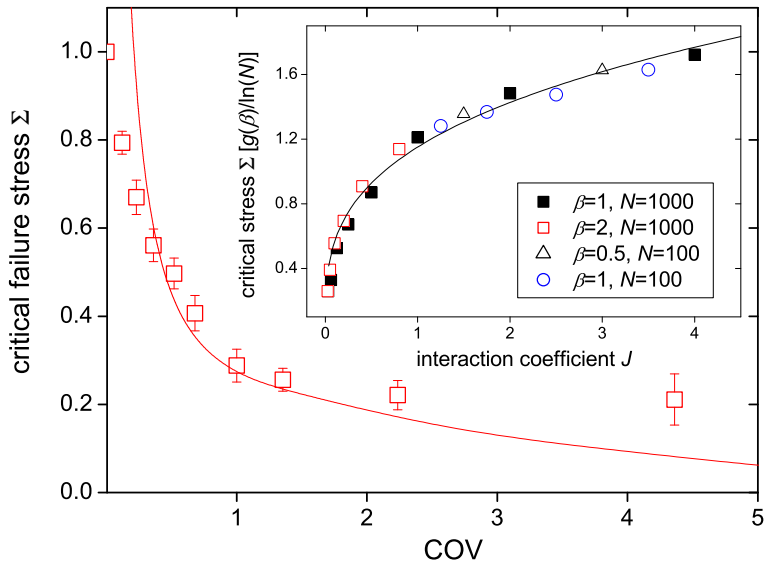}
  \caption{Main graph: mean failure strength vs. coefficient of variation, parameters: $J = 1, N=1000$; full line: theoretical
  fit (plot of $0.16g(\beta)$ vs. COV). Inset: mean failure strength, normalized by $g(\beta)/\ln(N)$, vs interaction
  coefficient $J$; full line: fit curve $1.15 J^{0.38}$.}
  \label{fig:strength}
\end{figure}

Figure \ref{fig:strength} shows the dependence of the mean failure strength on the value of the interaction coefficient and on the degree of disorder. As seen in the main graph, the decrease of strength with increasing disorder (increasing CoV/decreasing $\beta$) is  in the large-disorder regime well described by the theory  In the regime of small disorder, the theory predicts a diverging strength whereas the simulation for zero disorder yields a finite strength of 1. This is due to the finite value of the maximum stress that can be sustained by the interface; at this point the system fails by homogeneous yielding rather than by crack nucleation [for ideally brittle behavior the maximum stress would be infinite]. The effect can possibly be taken into account by accounting, in the derivation of the crack properties, for a finite process zone size as discussed by Bazant \cite{bazant98}. 

According to the theory, the characteristic failure stress is expected to scale as $\Sigma \propto g(\beta) \sqrt{2 J}/ \ln(N)$. The inset shows that failure stresses determined for different system sizes and Weibull exponents indeed collapse when scaled by $g(\beta)/\ln N$ and plotted vs. $J$, even though the fit function yields an exponent of $0.38$ rather than $0.5$. This discrepancy is attributed to logarithmic correction terms neglected in our derivation of the failure strength distribution. 

\section{Conclusions}

We have derived an analytical expression for the nucleation probability of one-dimensional cracks in a disordered interface between a thin film and a rigid substrate. The method we used is a variation of the method used by Arndt and Nattermann
\cite{arndt01} for deriving the nucleation probability of bulk cracks. Accordingly the obtained relations have strong structural similarities with those found in \cite{arndt01}. A main difference between the present problem and the one treated by Arndt and Nattermann resides, however, in the nature of the elastic interactions. Elasticity is essentially non-local for a bulk crack whereas, in the present case, the small thickness of the film leads to an effectively local stress re-distribution. From this point of view, our model may be considered a continuum version of local-load-sharing (LLS) fibre bundle models. Failure probability distributions for such models have been derived by several authors (see e.g \cite{phoenix00} and references therein). For certain types of disorder, such as a binary strength distribution, exact results can be obtained which confirm the existence of a logarithmic size effect. The present treatment, which uses extreme order statistics to obtain an expression for the size-dependent probability of system failure, is much more approximative. However, it has the advantage that the methods used can be applied to a broad class of local strength distributions with support in the interval $[0,\infty]$. Applying the present method to other local strength distributions (e.g. log-normal) leads to a similar structure of the failure probability distribution. The specific functional form of the distribution influences only the calculation of the function $g$ which characterizes the relation between the failure strength and the parameter governing the distribution width.

The present investigation focuses on one-dimensional cracks. This is a major restriction since in one dimension the crack shape does not depend on the disorder. In two dimensions, on the other hand, the crack shape may adjust to the strength fluctuations. In this case, the configuration space for crack nucleation becomes much more complicated as the problem can no longer be treated in terms of the crack length only, which in the present work serves as a single reaction coordinate. Instead, an investigation of crack nucleation and propagation in two dimensions requires consideration of the interplay between disorder and crack front elasticity, and a treatment of the ensuing crack front roughening. These aspects will be considered in a later publication.

\ack
Financial support of the European Commission under contract
NEST-2005-PATH-COM-043386 (NEST pathfinder programme TRIGS) 
as well as support of the ERC-Starting Grant MINATRAN (No. 211166) 
is gratefully acknowledged.

\end{document}